\documentclass[sigplan,10pt]{acmart}
\AtBeginDocument{%
    \providecommand\BibTeX{{%
          \normalfont B\kern-0.5em{\scshape i\kern-0.25em b}\kern-0.8em\TeX}}}
\usepackage{graphicx}
\usepackage{textcomp}
\usepackage[justification=centering]{caption}
\usepackage[justification=centering]{subcaption}
\usepackage{microtype}
\usepackage{listings}
    \usepackage{algorithm}
\usepackage{algpseudocode}
\usepackage{tikz}
\usetikzlibrary{shapes,fit}
\usepackage{url}
\usepackage{needspace}

\usepackage{url,afterpage,amsfonts}
\usepackage{enumitem,makecell}
\usepackage{booktabs}
\usepackage{times}
\usepackage{multirow}
\usepackage{color}
\usepackage{colortbl}
\usepackage{xcolor}
\usepackage{colortbl}
\usepackage{hyphenat}

\algtext*{EndWhile}
\algtext*{EndFor}
\algtext*{EndIf}

\AtBeginDocument{%
  \providecommand\BibTeX{{%
    \normalfont B\kern-0.5em{\scshape i\kern-0.25em b}\kern-0.8em\TeX}}}

\begin{document}
\acmYear{2021}\copyrightyear{2021}
\setcopyright{acmcopyright}
\acmConference[PPoPP '21]{26th ACM SIGPLAN Symposium on Principles and Practice of Parallel Programming}{February 27--March 3, 2021}{Virtual Event, Republic of Korea}
\acmBooktitle{26th ACM SIGPLAN Symposium on Principles and Practice of Parallel Programming (PPoPP '21), February 27--March 3, 2021, Virtual Event, Republic of Korea}
\acmPrice{15.00}
\acmDOI{10.1145/3437801.3441611}
\acmISBN{978-1-4503-8294-6/21/02}
\title{BiPart: A Parallel and Deterministic Multilevel Hypergraph Partitioner}

\author{Sepideh Maleki\texorpdfstring{$^{*}$}{}}
\affiliation{
  \institution{The University of Texas at Austin}            
}
\email{smaleki@cs.utexas.edu}          

\author{Udit Agarwal\texorpdfstring{$^{*}$}{}}
\affiliation{
  \institution{The University of Texas at Austin}            
}
\email{udit@utexas.edu}          
\author{Martin Burtscher}
\affiliation{
  \institution{Texas State University}            
}
\email{burtscher@txstate.edu}          
\author{Keshav Pingali}
\affiliation{
  \institution{The University of Texas at Austin}            
}
\email{pingali@cs.utexas.edu}          

\renewcommand{\shortauthors}{Maleki, Agarwal, Burtscher, and Pingali}

\begin{abstract}
Hypergraph partitioning is used in many problem domains including VLSI design, linear algebra, Boolean satisfiability, and data mining. Most versions of this problem are NP-complete or NP-hard, so practical hypergraph partitioners generate approximate partitioning solutions for all but the smallest inputs. One way to speed up hypergraph partitioners is to exploit parallelism. However, existing parallel hypergraph partitioners are not deterministic, which is considered unacceptable in domains like VLSI design where the same partitions must be produced every time a given hypergraph is partitioned.

In this paper, we describe BiPart, the first deterministic, parallel hypergraph partitioner. Experimental results show that BiPart outperforms state-of-the-art hypergraph partitioners in runtime and partition quality while generating partitions deterministically.

\end{abstract}

\maketitle

\let\thefootnote\relax\footnote{$^*$ Both authors contributed equally.}

\section{Introduction}
\label{sec:intro}


A \emph{hypergraph} is a generalization of a graph in which an edge can connect any number of nodes. Formally, a hypergraph is a tuple $(V,E)$ where $V$ is the set of \emph{nodes} and $E$ is a set of nonempty subsets of $V$ called {\em hyperedges}. Graphs are a special case of hypergraphs in which each hyperedge connects exactly two nodes~\cite{berge1973}.

\begin{figure}[H]
    \begin{subfigure}[b]{0.45\textwidth}
      \includegraphics[width=\textwidth,page=1]{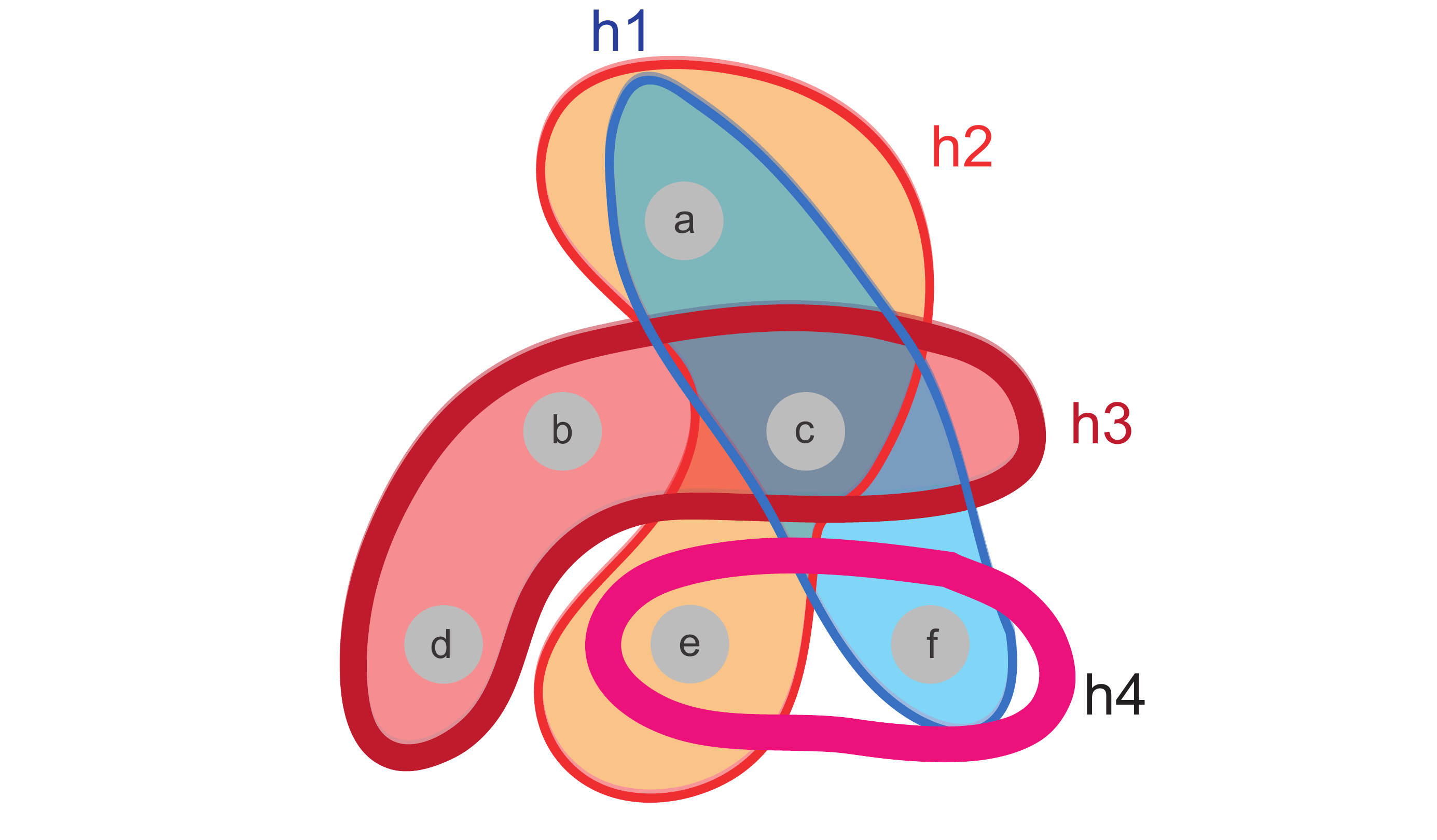}
      \caption{A hypergraph}
      \label{fig:hg_example1}
    \end{subfigure}
    \begin{subfigure}[b]{0.45\textwidth}
      \includegraphics[width=\textwidth,page=2]{hypergraphs.pdf}
      \caption{Bipartite graph representation}
      \label{fig:hg_example2}
    \end{subfigure}
  \caption{Example hypergraph and the corresponding bipartite graph representation} \label{fig:hg_example}
  \end{figure}

Figure~\ref{fig:hg_example1} shows a hypergraph with 6 nodes and 4 hyperedges. The hyperedges are shown as colored shapes around nodes. The \emph{degree} of a hyperedge is the number of nodes it connects. In the figure, hyperedge h1 connects nodes  \textit{a},  \textit{c}, and  \textit{f}, and it has a degree of three.

Hypergraphs arise in many application domains. In VLSI design, circuits are often modeled as hypergraphs; nodes in the hypergraph represent the pins of the circuit and hyperedges represent wires from the output pin of a gate to the input pins of other gates~\cite{Caldwell99}. In Boolean satisfiability, a Boolean formula can be represented as a hypergraph in which nodes represent clauses and hyperedges represent the occurrences of a given literal in these clauses. Hypergraphs are also used to model data-center networks~\cite{datacenter}, optimize storage sharding in distributed databases~\cite{facebook}, and minimize the number of transactions in data centers with distributed data~\cite{disdatacenter}.

\subsection{Hypergraph partitioning}

In many of these applications, it is necessary to partition the hypergraph into a given number of subgraphs. For example, one of the key steps in VLSI design, called placement, assigns a location on the die to each gate. Good algorithms for placement must balance competing goals: to avoid hotspots on the chip, it is important to spread out circuit components across the entire die but this may increase interconnect wire lengths, reducing the rate at which the chip can be clocked. This problem is often solved using hypergraph partitioning~\cite{Caldwell99}. Hypergraph partitioning is also used to optimize logic synthesis~\cite{LSOracle}, sparse-matrix vector multiplication~\cite{patoh}, and storage sharding~\cite{facebook}.

Formally, the k-way hypergraph partitioning problem is defined as follows. Given a  hypergraph G = (V, E), the number of partitions to be created ($k \geq 2$), and an \emph{imbalance parameter} ($\epsilon \geq 0$), a k-way partition $P = \{V_1,V_2...,V_k\}$
is said to be {\em balanced} if it satisfies the constraint $|V_i| \leq (1+\epsilon)(|V|/k)$. Given a partition of the nodes, each hyperedge is assigned a {\em penalty} equal to one less than the number of partitions that it spans; intuitively, a hyperedge whose nodes are all in a single partition has zero penalty, and the penalty increases as the number of partitions spanned by the hyperedge increases. The penalty for the partition is defined to be the sum of the penalties of all hyperedges. Formally, $cut(G,P) =  \sum_{e} (\lambda_e(G,P) - 1)$, where $\lambda_e(G,P)$ is the number of partitions that hyperedge $e$ spans. The goal of hypergraph partitioning is to find a balanced partition that has a minimal cut. In some applications, hyperedges have weights, in which case the contribution to $cut(G,P)$ from each hyperedge $e$ in the definition above is multiplied by the weight of $e$. 

Many partitioners produce two partitions (often called \emph{bipartitions}), and this step is repeated recursively to obtain the required number of partitions.

Although graph partitioners have been studied extensively in the literature~\cite{Fiedler73,KL,Karypis,multilevel}, there has been relatively
little work on hypergraph partitioning. In principle, graph partitioners can be used for hypergraph partitioning by converting a hypergraph into a graph, which can be accomplished by replacing each hyperedge with a clique of edges connecting the same nodes. However, this transformation increases the memory requirements of the partitioner substantially if there are many large hyperedges and may lead to poor-quality partitions~\cite{Caldwell99}. Therefore, it is often better to treat hypergraphs separately from graphs. One way to represent a hypergraph $H$ concretely is to use a bipartite graph $G$ as shown in Figure~\ref{fig:hg_example}. In $G$, one set of nodes represents the hyperedges in $H$, the other set of nodes represents the nodes in $H$, and an edge $(u,v)$ in \emph{G} is used to represent the fact that, in the hypergraph, the hyperedge represented by $u$ contains the node represented by $v$.

An ideal hypergraph partitioner has three properties.

\begin{enumerate}
\item The partitioner should be capable of partitioning \emph{large} hypergraphs with millions of nodes and hyperedges, producing high-quality partitions within a few seconds.

\item In some domains like VLSI circuit design, the partitioner must be \emph{deterministic}; {\em i.e.}, for a given hypergraph, it must produce the same partitions every time it is run \emph{even if the number of threads is changed from run to run}. For example, the manual post-processing in VLSI design after partitioning optimizes the placement of the cells within each partition. Many placement tools can do efficient placement only for standard cells, and if non-standard cells are used, the placement may need to be optimized manually. Deterministic partitioning is essential to avoid having to redo the placement.

\item Since hypergraph partitioners are based on heuristics, they have \emph{parameters} whose optimal values may depend on the hypergraph to be partitioned. Hypergraph partitioners should permit design-space exploration of these parameters by sophisticated users.
\end{enumerate}

Most variations of graph and hypergraph partitioning are either NP-complete or NP-hard~\cite{nphard}, so heuristic methods are used in practice to find good solutions in reasonable time. Prior work in this area is surveyed in Section~\ref{sec:bg}~\cite{Karypis,hmetis,phmetis,patoh,zoltan,pothen90,kahypar,smaleki}.

In our experience, existing partitioners lack one or more of the desirable properties listed above. Many high-quality hypergraph partitioners like HMetis~\cite{hmetis}, PaToH~\cite{patoh}, and KaHyPar~\cite{kahypar} are serial programs. For some of the hypergraphs in our test suite, these partitioners either run out of memory or time out after an hour, as described in Section~\ref{sec:exp}.

Parallel hypergraph partitioners like Zoltan~\cite{zoltan} and the Social Hash Partitioner from Facebook~\cite{facebook} can handle all hypergraphs in our test suite, but they are nondeterministic (we have observed that, for a hypergraph with 9 million nodes, the edge-cut in the output of Zoltan can vary by more than 70\% from run to run when using different numbers of cores). It is important to note that this nondeterminism does not arise from incorrect synchronization of parallel reads and writes but from \emph{under-specification} in the program; for example, the program may make a random selection from a set, and although it is correct to choose any element of that set,
different choices may produce different outputs. Parallel programming systems may exploit such {\em don't-care nondeterminism} to improve parallel  performance~\cite{pingali11}, but parallel partitioners with don't-care nondeterminism will violate the second requirement listed above.

\subsection{BiPart}

These limitations led us to design and implement BiPart, a parallel, deterministic hypergraph partitioner that can partition all the hypergraphs in our test suite in just a few seconds. This paper makes the following contributions.
\begin{itemize}
  \item We describe BiPart, an open-source framework for parallel, \emph{deterministic} hypergraph partitioning.
  \item We describe \emph{application-level} mechanisms that ensure that partitioning is deterministic even though the runtime exploits don't-care nondeterminism for performance.
  \item We describe a novel strategy for parallelizing multiway partitioning.
  \item We show experimentally that BiPart outperforms existing hypergraph partitioners
  in either partition quality or running time, and usually outperforms them in both dimensions.
\end{itemize}

The rest of the paper is organized as follows.  Section~\ref{sec:related_work} describes background and related work on hypergraph partitioning.
Section~\ref{sec:hypar} describes BiPart, our deterministic parallel hypergraph partitioner. Section~\ref{sec:exp} presents and analyzes the experimental results on a shared-memory NUMA machine. Section~\ref{sec:conclusions} concludes the paper.

\section{Prior Work on Graph and Hypergraph Partitioning}
\label{sec:bg}
\label{sec:related_work}

There is a large body of work on graph and hypergraph partitioners,
so we discuss only the most closely related work in this section.
It is useful to divide partitioners into \emph{geometry-based partitioners}
(Sec. 2.1) and \emph{topology-based partitioners}
(Sec. 2.2). \emph{Multilevel partitioning}, discussed in Sec. 2.3, adds a different dimension to partitioning. BiPart uses a topology-based multilevel partitioning approach. 

\subsection{Geometry-based Partitioning}
\label{sec:geometric}
\label{sec:spectral}
In some domains such as finite elements, the nodes of the graph are points in a metric space such as $\mathbb{R}\textsuperscript{d}$, so we can compute the distance between
two nodes. The geometric notion of proximity of nodes can be used to partition the graph using techniques like k-nearest-neighbors (KNN)~\cite{mitchell1997machine}.
A sophisticated geometric partitioner was introduced by
Miller, Teng, and Vavasis~\cite{vavasis}.
This partitioner stereographically projects nodes from $\mathbb{R}^d$ to a sphere in
$\mathbb{R}^{d{+}1}$. The sphere is bisected by a suitable great circle, creating the partitions,
and the nodes are projected back to $\mathbb{R}^d$ to obtain the partitions.

When there is no geometry associated with the nodes of a graph,
\emph{embedding techniques} can be used to map nodes to points in $\mathbb{R}^d$ in ways
that try to preserve proximity of nodes in the graph; geometry-based
partitioners can then be used to partition the embedded graph.

One powerful but expensive embedding technique is based on computing
the Fiedler vector of the Laplacian matrix of a graph~\cite{Fiedler73}.
The Fiedler vector is the eigenvector corresponding to the second smallest
eigenvalue of the Laplacian matrix. The Fiedler vector
is a real vector (it can be considered an embedding of the nodes in
$\mathbb{R}^1$) and the signs of its entries can be used to determine
how to partition the graph. Several \emph{spectral partitioners} based on this idea were implemented
and studied in the mid-90's~\cite{pothen90}. They can produce good graph
partitions since they take a global view of the graph, but they are not
practical for large graphs.

Heuristic embedding techniques known as {\em node2vec} or
{\em DeepWalk} are currently receiving a lot of attention in the machine-learning
community~\cite{node2Vec,DeepWalk}. These techniques are based on random walks in
the graph to estimate proximity among nodes, and these estimates are
used to compute the embedding. Techniques like stochastic gradient descent (SGD)
are employed to iteratively improve the embedding. 

Unfortunately, all embedding techniques we know of are computationally intensive
so they cannot be used for large graphs without geometry
if partitioning is to be done quickly.

\subsection{Topology-based Partitioning}
\label{sec:local}
In contrast to geometry-based partitioners, topology-based partitioners
work only with the connectivity of nodes in the graph or hypergraph. These partitioners
generally start with some heuristically chosen partitioning and then apply \emph{local} refinements
to improve the balance or the edge cut until a termination condition is reached.

Kernighan and Lin invented one of the first practical graph partitioners.
An initial bipartition of the graph is obtained using a technique such as
a breadth-first traversal of the graph, starting from an arbitrary
node and terminating when half the nodes have been touched.
Given such a partitioning of the graph that is well balanced,
the algorithm (usually called the \emph{KL} algorithm) attempts to reduce the cut by
swapping pairs of nodes between the partitions until a termination criterion
is met~\cite{KL}.

Fiduccia and Mattheyses generalized this algorithm to
hypergraphs (their algorithm is usually referred to as the \emph{FM} algorithm)~\cite{FM}. It starts by computing the gain values for each node, where gain refers to the change in the edge cut if a node were moved to the other partition. The algorithm executes in rounds; in each round,
a subset of nodes are moved from their current partition to the other partition. A greedy algorithm is used to identify this subset:
the node with the highest gain value is selected to be moved,
the gain values of its neighbors are updated accordingly,
and the process is repeated with the remaining unmoved nodes until
all nodes are moved exactly once. At the end of every round, the algorithm picks the maximal prefix of these moves that results in the highest gain and moves the rest of the nodes back to their original partition. The overall algorithm terminates when no gain is achieved in the current round.

Experimental studies show that the quality of the partitions produced
by these techniques depends critically on the quality of the initial
partition. Intuitively, these algorithms perform local optimization,
so they can improve the quality of a good initial partition but they
cannot find a high quality partition if the initial partition is poor,
since this requires global optimization.

\subsection{Multilevel Graph Partitioning}
\label{sec:multilevel}
Multilevel partitioning techniques attempt to \emph{circumvent} the
limitations of the algorithms described above rather than replace them
with an entirely new algorithm. This approach was
first explored for graphs~\cite{metis1, metis2, Karypis}
 and later extended to hypergraphs in
the HMetis partitioner~\cite{hmetis}. Since every graph is a hypergraph, we use the term {\em hypergraph} to include graphs in the rest of the paper.

Multilevel hypergraph partitioning consists of three phases: \emph{coarsening}, \emph{initial partitioning}, and \emph{refinement}.

\begin{itemize}
  \item {\em Coarsening}: For a given hypergraph $G_f$, a coarsened hypergraph $G_c$ is created by merging pairs of nodes in $G_f$. We call $G_c$ the coarsened hypergraph and $G_f$ the fine-grained hypergraph. This process can be applied recursively to the coarsened hypergraph, creating a chain of hypergraphs in which the first hypergraph is the initial hypergraph and the final hypergraph is a coarsened hypergraph that meets some termination criterion ({\em e.g.}, its size is below some threshold).

  \item {\em Initial partitioning}: The coarsest hypergraph is partitioned
  using any of the techniques discussed in Sections~\ref{sec:geometric} and \ref{sec:local}.

  \item {\em Refinement}: For each pair $G_c$ and $G_f$, the
  partitioning of $G_c$ is projected onto $G_f$ and then refined,
  starting from the most coarsened hypergraph and finishing
  with the input hypergraph.
\end{itemize}

Various heuristics have been implemented for these three phases. For example, heavy-edge matching,
where a node tries to merge with the neighbor with which it shares the heaviest weighted
edge, is widely used in coarsening~\cite{Karypis}. Techniques frequently used in refinement
include swapping pairs of nodes from different partitions, as in the KL algorithm,
or moving nodes from one partition to another, as in the FM algorithm.
Most of these heuristics were designed for sequential implementations so they cannot
be used directly in a parallel implementation.

\subsection{Parallel Hypergraph Partitioning}
\label{sec:bg_parallel}
\label{sec:bg_determinisim}
Hypergraph partitioners should be parallelized to prevent them from becoming
the performance bottleneck in hypergraph processing. Zoltan~\cite{zoltan} and Parkway~\cite{parkway}
are parallel hypergraph partitioners based on the multilevel scheme.
HyperSwap~\cite{hyperswap} is a distributed algorithm that partitions
hyperedges instead of nodes. The Social Hash partitioner~\cite{facebook}
is another distributed partitioner for balanced k-way hypergraph partitioning.

One disadvantage of these parallel hypergraph partitioners is
that their output is nondeterministic. For example, in
the coarsening phase, it may be desirable to merge a given node
$V_1$ with either node $V_2$ or node $V_3$. In a parallel
implementation, slight variations in the internal timing between
executions may result in choosing different nodes
for merging, producing different partitions of the same input graph.
However, many applications require deterministic partitioning,
as discussed in Section~\ref{sec:intro}.

\subsection{Ensuring determinism}

The problem of ensuring deterministic execution of parallel programs with don't-care nondeterminism has been studied at many abstraction levels. At the systems level,
there has been a lot of work on ensuring that parallel threads communicate in a deterministic manner~\cite{devietti2011rcdc, hower2011calvin, kaler2016executing}. For many programs,
this ensures deterministic output if the program is executed on the \emph{same}
number of threads in every run. However, it does not address our requirement that the output of the partitioner must be the same even if the number of threads on which it executes is \emph{different} in different runs. Moreover, these solutions usually result in a substantial slowdown~\cite{nguyen:2014,devietti2011rcdc}. 

For nested task-parallel programs, an approach called {\em internal determinism} has been proposed to ensure that the program is executed in deterministic steps, thereby ensuring that the output is deterministic as well~\cite{blelloch2012internally}. The Galois system solves the determinism problem in its task scheduler~\cite{nguyen:2014}, which finds independent sets of tasks in an implicitly constructed interference graph. To guarantee a deterministic schedule, the independent set must be selected in a deterministic fashion. This is achieved without building an explicit interference graph.  The neighborhood items of a task are marked with the task ID, and ownership of neighborhood items with lower ID values are stolen during the marking process. An independent set is then constructed by selecting the tasks whose neighborhood locations are all marked with their own ID values.

Both these solutions guarantee that the output does not depend on the number of threads used to execute the program. However, our experiments showed that these generic, application-agnostic solutions are too heavyweight for use in hypergraph partitioning, so we devised a lightweight {\em application-specific} technique for ensuring determinism with substantially less overhead as described in Section~\ref{sec:hypar}.

\section{BiPart: A Deterministic Parallel Hypergraph Partitioner}
\label{sec:hypar}

This sections describes BiPart, our deterministic parallel multilevel hypergraph partitioner. BiPart produces a bipartition of the hypergraph, and it is used recursively on these partitions to produce the desired number of partitions.

\subsection{Coarsening}
\label{sec:hypar_coarsen}

\begin{figure*}
\begin{center}

\includegraphics[scale=0.3]{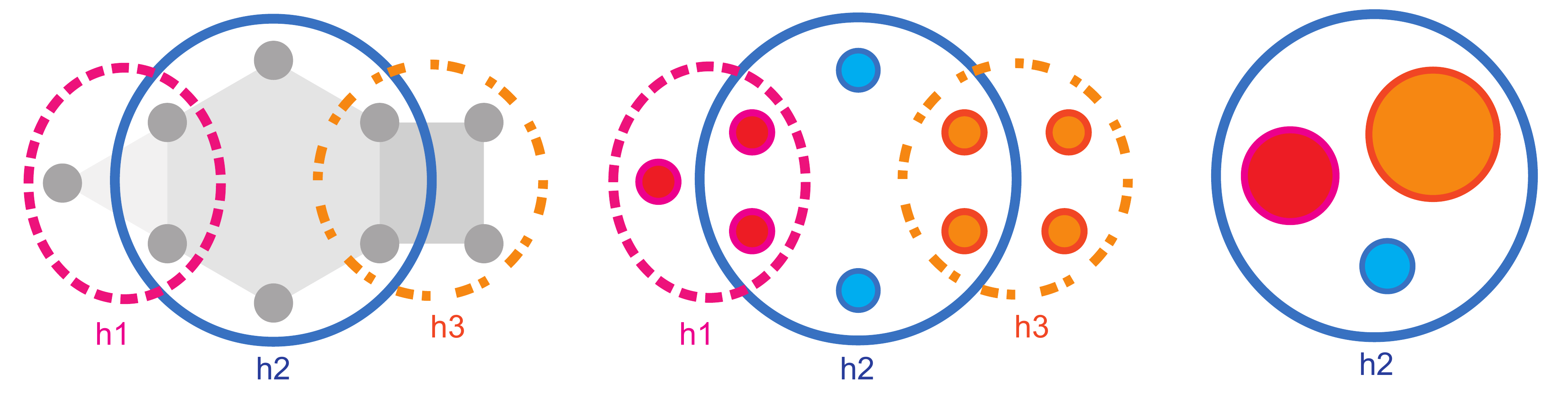}
  \caption{Multi-node coarsening: (a) a hypergraph with 3 hyperedges, h1, h2, and h3 (left). 
  (b) multi-node matching matches nodes within a hypergraph (center). (c) merging  matched nodes coarsen hypergraph (right). } \label{fig:coarse}
\end{center}
\end{figure*}

\begin{algorithm}
\small
\caption{Parallel Matching Policy}\label{matching}
\begin{algorithmic}[1]
\Statex \textbf{Input:} $fineGraph$, $policy$; 
\Statex \textbf{/* Initialize node priorities */}
\ForAll{nodes $node \in  fineGraph$ \textbf{in parallel}}	\label{alg0:for0}
    \State $node.priority \leftarrow \infty$
    \State $node.rand \leftarrow \infty$
    \State $node.hedgeid \leftarrow \infty$
\EndFor	\label{alg0:endFor0}
\Statex \textbf{/* Assign priorities based on the policy (e.g. low degree hyperedges) */}
\ForAll{hyperedges $hedge \in  fineGraph$ \textbf{in parallel} }	\label{alg0:for1}
 	\State $hedge.priority \leftarrow degree(hedge)$    \label{alg0:hedgePriority}
  	\State $hedge.rand \leftarrow hash(hedge.id)$   \label{alg0:hedgeRand}
  	\For{$node \in hedge$}  \label{alg0:for1:for}
		  \State $node.priority \leftarrow atomicMin(node.priority,$ \\ \hspace{1.2in} $ hedge.priority)$
	 \EndFor
 \EndFor	\label{alg0:endFor1}
  \Statex \textbf{/* Assign a second priority (hash of hedge id) */}
  \ForAll{hyperedges $hedge \in  fineGraph$ \textbf{in parallel} }	\label{alg0:for2}
    \For{node $\in$ hedge}
      \If{$hedge.priority == node.priority$}
		  \State $node.rand \leftarrow atomicMin (node.rand, hedge.rand)$
	  \EndIf
   \EndFor 
\EndFor	\label{alg0:endFor2}
\Statex \textbf{/* Assign each node to its incident hyperedge with highest priority */}
\ForAll{hyperedges $hedge \in  fineGraph$ \textbf{in parallel} }	\label{alg0:for3}
  \For{node $\in$ hedge}
      \If{$hedge.rand == node.rand$}
		  \State $node.hedgeid \leftarrow atomicMin(node.hedgeid,$ \\ \hspace{1.4in} $hedge.id)$
	  \EndIf
  \EndFor 
\EndFor	\label{alg0:endFor3}
\end{algorithmic}
\end{algorithm}

\begin{algorithm}
\small
\caption{Parallel Coarsening}\label{euclid}
\begin{algorithmic}[1]
\Statex \textbf{Input:} $fineGraph$, $policy$; \textbf{Output:} $coarseGraph$
\State Find a multi-node matching $M$ of $fineGraph$ using Algorithm~\ref{matching}  \label{alg1:matching}
\Statex \textbf{/* Merge nodes of the finer graph */}
  \ForAll{hyperedges $hedge \in  fineGraph$ \textbf{in parallel}}	\label{alg1:for1}
    \State $S$: Set of nodes in $hedge$ that are matched
    \If{$|S| > 1$}
      \State $N \leftarrow $ Merge nodes in $S$
	  \ForAll{$node \in S$}	\label{alg1:for1:for1}
	    \State $parent(node) \leftarrow N$
	  \EndFor	\label{alg1:for1:endFor1}
    \EndIf
  \EndFor	\label{alg1:endFor1}
  \Statex \textbf{/* Merge singleton nodes with an already merged node */}
  \ForAll{hyperedges $hedge \in  fineGraph$ \textbf{in parallel} }	\label{alg1:for2}
      	\State $S$: Set of nodes in $hedge$ that are matched \label{alg1:S2}
	  \If{$|S| = 1$ and there exist an already merged node $v \in hedge$ \label{alg2:Sing2}}
		  \State $v$: Merged nodes in hedge with smallest weight \label{alg1:let}
		  \State Merge node $u$ in $S$ with $v$ 
		  \State $parent(u) \leftarrow parent(v)$
		  \Statex \textbf{/* Self merge singleton nodes */}
		  \ElsIf{$|S| = 1$}
		  \State $parent(S) \leftarrow S$ 
	  \EndIf
  \EndFor	\label{alg1:endFor2}
  \Statex \textbf{/* Create hyperedges in the coarsened graph */}
  \ForAll{hyperedges $hedge \in fineGraph$ \textbf{in parallel} }	\label{alg1:for3}
	  \State $parents \leftarrow \emptyset$
  	  \ForAll {$node \in hedge$}
		  \If{$parent(node) \notin parents$}
		     \State $N \leftarrow coarseGraph.createNode(parent(node)) $
		  	 \State $parents.add(parent(node))$
		  	 \EndIf
  	  \EndFor
	  \If{$|parents| > 1$}
		  \State $E \leftarrow coarseGraph.createHyperedge()$
		  \State $parent(hedge) \leftarrow E$
		  \ForAll{$node \in parents$}
			  \State $includeNodeInEdge(E, node)$
		  \EndFor
	  \EndIf
  \EndFor
\end{algorithmic}	\label{algCoarsening}
\end{algorithm}

The goal of coarsening is to create a series of smaller hypergraphs until a small enough hypergraph is obtained that can be partitioned using a simple heuristic. Intuitively, coarsening finds nodes that should be assigned to the same partition and merges them to obtain a smaller hypergraph. However, it is important to reduce the size of hyperedges as well since this enables the subsequent refinement phase to be more effective (FM and related algorithms are most effective with small hyperedges).

Coarsening can be described using the idea of \emph{matchings} from graph theory~\cite{berge1973}.

\begin{description}
\item [Hyperedge matching:] A \textit{hyperedge matching} of a hypergraph $H$ is an independent set of hyperedges such that no two of them have a node in common.
  In Figure~\ref{fig:hg_example}, \{h3, h4\} is a hyperedge matching. 

\item[Node matching:] A \textit{node matching} of a hypergraph $H$
is a set of node pairs $(u,v)$, where $u$ and $v$ belong to the same hyperedge such that no two pairs have a node in common. In Figure~\ref{fig:hg_example}, \{(a,e), (b,c)\} is a node matching.

\item[Multi-node matching:] BiPart uses a modified version of node matching called \textit{multi-node matching}, where instead of node pairs we have a partition of the nodes of $H$ such that each node set in the partition contains nodes belonging to one hyperedge. In Figure~\ref{fig:hg_example}, \{(a,e), (b,c,d), (f)\} is a multi-node matching.
\end{description}

Coarsening can be performed by contracting nodes or hyperedges. In the \textit{node coarsening} scheme, 
a node matching is first computed and the
nodes in each node pair in the matching are then merged together.
\textit{Hyperedge coarsening} computes a hyperedge matching, and all nodes connected by a hyperedge in this matching are merged to form a single node in the coarsened hypergraph.

In contrast, BiPart uses multi-node matching, which has advantages over both node coarsening and hyperedge coarsening. A hyperedge disappears from a coarsened graph only after all its member nodes are merged into one node. In node coarsening, the number of hyperedges may stay roughly the same even after merging the nodes in the matching. Similarly in hyperedge coarsening, the hyperedge matching may have a very small size and may result in only a small reduction in the size of the hypergraph. The coarsening phase in BiPart consists of two parts: finding a multi-node matching and the coarsening algorithm. 
\begin{table}
\small
\caption{Matching policies for multi-node matching}
\label{t:match_hyperedge}
\begin{tabular}{|c|c|}
\hline
\textbf{Policy}&\textbf{Policy Description} \\
\hline
	\textit{LDH}& \textit{Hyperedges with lower degree have higher priority} \\
\hline
	\textit{HDH}& \textit{Hyperedges with higher degree have higher priority} \\
\hline
	\textit{LWD}& \textit{Lower weight hyperedges have higher priority} \\
\hline
	\textit{HWD}& \textit{Higher weight hyperedges have higher priority} \\
\hline
\textit{RAND}& \textit{Priority assigned by a deterministic hash of ID value} \\
\hline
\end{tabular}
\end{table}

\subsubsection{Finding a multi-node matching}	\label{sec:matching}
Algorithm~\ref{alg1:matching} lists the pseudocode of multi-node matching. BiPart computes a multi-node matching in the following way: 
First, every hyperedge is assigned a priority based on a matching policy and a deterministic hash of its ID value (Lines~\ref{alg0:hedgePriority} -~\ref{alg0:hedgeRand}).
The matching policy can be based on the degree of the hyperedge, weight, etc. Table ~\ref{t:match_hyperedge} lists the available matching policies for BiPart. 
Every node is then assigned a piority value, which is the minimum across all its incident hyperedges (Lines~\ref{alg0:for1:for}-\ref{alg0:endFor1}).
In case many hyperedges have identical degrees, every node
is assigned a second priority value (Lines~\ref{alg0:for2}-\ref{alg0:endFor2}) to reduce contention.
Finally, every node matches itself to one of its incident hyperedges with the highest priority, 
e.g., the hyperedge with the
lowest degree and with the lowest hashed value (in case the hyperedges have the same degree) (Lines~\ref{alg0:for3}-~\ref{alg0:endFor3}).
The nodes that are matched to the same hyperedge are then grouped together, 
resulting in a deterministic multi-node matching.

\subsubsection{Coarsening Algorithm}
Algorithm~\ref{euclid} lists the pseudocode of a single phase of the coarsening algorithm used in BiPart. We perform this step repeatedly for at most \textit{coarseTo} iterations. Coarsening consists of two steps. First, BiPart merges all the nodes that are matched to the same hyperedge into a single node in the coarsened graph (Lines~\ref{alg1:matching}-\ref{alg1:endFor1}). For optimization purposes, we ignore the singleton sets and BiPart instead merges nodes in such sets with a neighbor node that has been merged in the previous step (Lines~\ref{alg2:Sing2}-\ref{alg1:endFor2}).

Figure~\ref{fig:coarse} illustrates this on a hypergraph with nine nodes and three hyperedges h1, h2, and h3. In the first step, BiPart performs multi-node matching (priority is with the low degree hyperedges (LDH)), Figure~\ref{fig:coarse} (center). Figure~\ref{fig:coarse} (right) shows the result of this matching. The nodes in each of the disjoint sets in the matching are merged into a single node. Note that, since all nodes of hyperedges h1 and h3 are merged to a single node, we can remove those hyperedges and only h2 remains in the hypergraph.

\subsubsection{Ensuring Determinism.}
Step~\ref{alg1:matching} in the coarsening phase, which finds a multi-node matching of the hypergraph, is a potential source of nondeterminism. The approach presented in Section~\ref{sec:matching} yields a deterministic multi-node matching. This matching is used to coarsen the graph deterministically.


\subsection{Initial Partitioning}

\begin{algorithm}
\small
\caption{Initial Partitioning Algorithm}
\begin{algorithmic}[1]
\Statex \textbf{Input:} coarsest graph $G_x = (V_x,E_x)$
\Statex \textbf{Output:} Partitions $P_0$ and $P_1$.
\State $P_0$ = \{\}; $P_1$ = $V_x$
\State n = $|V_x|$
\State Compute move gain values for nodes in $P_1$ using Algorithm~\ref{algMoveGain}
\While{$|P_0| < |P_1|$}	\label{startWhile}
	\State Pick $\sqrt{n}$ nodes from $P_1$ with highest gain values (break ties using node ID) and move them to $P_0$ \textbf{in parallel}	\label{move}
	\State Re-compute move gain values for nodes in $P_1$ using Algorithm~\ref{algMoveGain}	\label{algInitialLarge:recompute}
\EndWhile	\label{endWhile}
\end{algorithmic}  \label{algInitialLarge}
\end{algorithm}

The goal of this step is to obtain a good bipartition of the coarsest graph.
There are many ways to accomplish this but the key idea in most
algorithms is to maintain two sets of nodes $P_0$ and $P_1$ where
$P_0$ and $P_1$ contain the nodes assigned to partitions $0$ and $1$, respectively. 
Iteratively, some nodes from $P_1$ are selected and
moved to $P_0$ (assuming $P_0$ is smaller than $P_1$) until the balance condition is met.

The selection of nodes can be implemented in many ways. A simple approach is to do a breadth-first search (BFS) of the graph starting from some arbitrary vertex. In this approach, nodes on the BFS frontier are selected at each step for inclusion in the partition. The greedy graph-growing partitioning algorithm (GGGP) used in Metis maintains
gain values for every node $v$ in $P_1$ ({\em i.e.}, the decrease in the edge cut
if $v$ is moved to the growing partition) and it always picks the node with
the highest gain at each step and updates the gain values of the remaining nodes
in $P_1$. 
However, this GGGP approach is inherently serial.

Instead, BiPart uses a more parallel approach to obtain an initial partition.
The approach used in BiPart is the following. Like GGGP, we maintain gain values
for nodes in $P_1$, but we pick the top $\sqrt{n}$
nodes with the highest gain values in each step and move them to $P_0$ (here $n$ denotes
the number of nodes in the coarsest graph). We then re-compute the gain values of all
nodes in $P_1$.
This gives us a good parallel algorithm for computing the initial partition.
Algorithm~\ref{algInitialLarge} lists the pseudocode.

Algorithm~\ref{algMoveGain} describes the pseudocode for computing move gain values. It is based on the approach used in the FM algorithm~\cite{FM}.

\begin{algorithm}
\small
\caption{Compute Move-Gain Values}
\textbf{Input:} Graph $G = (V,E)$, $P_0$ and $P_1$ are the two partitions
	\begin{algorithmic}[1]
		\State Initialize $Gain(u)$ to $0$ for all $u \in V$ in parallel \label{algMove:init}
		\For{all hyperedges $hedge \in E$   \textbf{in parallel}}	\label{algMoveGain:for1}
			\State $n_0$ $\leftarrow$ number of nodes in $P_0 \cap hedge$
			\State $n_1 \leftarrow$ number of nodes in $P_1 \cap hedge$
			\For{$u \in hedge$}
				\State $i \leftarrow$ partition of $u$
				\If{$n_i == 1$}	\Comment{$u$ is the only node from $P_i$ in $hedge$}
					\State $Gain(u) \leftarrow Gain(u) + 1$
				\ElsIf{$n_i == |hedge|$}	\Comment{all nodes are in $P_i$}
					\State $Gain(u) \leftarrow Gain(u) - 1$
				\EndIf
			\EndFor
		\EndFor	\label{algMoveGain:endFor1}
	\end{algorithmic}  \label{algMoveGain}
\end{algorithm}

\subsubsection{Ensuring Determinism.}
In the initial partitioning phase, nondeterminism may be present in Line 5 of Algorithm ~\ref{algInitialLarge} where we need to pick a node $v$ with highest gain value and
there are multiple nodes with the same highest gain. To ensure determinism, BiPart again breaks ties using node IDs.

\subsection{Refinement Phase}

The third phase of the overall partitioning algorithm is the refinement phase.
The goal of this phase is to improve on the bipartition obtained from the initial partitioning. This phase runs a refinement algorithm on the sequence of graphs obtained during the coarsening phase, starting from the coarsest graph and terminating at the original input graph. The FM refinement algorithm described in Section~\ref{sec:local} is inherently serial and cannot be used for large graphs as it is, since it needs to make individual moves for every node in every pass. Our refinement algorithm, in contrast, makes parallel node moves, thus speeding up the process. However, this approach may result in a poor edge cut since it does not choose the best prefix of moves, unlike the FM algorithm. We address this issue by ensuring that we only move nodes with high or positive gain values.

Another major difference in our refinement algorithm is that we do not consider the weight of the nodes when making these moves. This helps in speeding up the algorithm but may result in an unbalanced partition. We resolve this possible issue by running a separate balancing algorithm after the refinement. Algorithm~\ref{algRefine} provides the pseudocode of our refinement approach. The input to the algorithm is an integer {\em iter} that specifies the number of rounds of refinement to be performed; a larger number of rounds may improve partition quality at the cost of extra running time.

\begin{algorithm}
\small
\caption{Refinement Algorithm}
\begin{algorithmic}[1]
\Statex \textbf{Input:} $iter$: refinement iterations; Partitions $P_0$ and $P_1$
\State \textbf{Initialization:} Project bipartition from coarsened graph	\label{Refine:project}
\For{$iter$ iterations}	\label{Refine:for}
	\State \hspace{0.05in} Compute move gain values for all nodes using Algo~\ref{algMoveGain} \label{Refine:compute}
	\State \hspace{0.05in} $L_0 \leftarrow$ nodes in $P_0$ with gain value $\geq 0$ \label{Refine:filterP1}
	\State \hspace{0.05in} $L_1 \leftarrow$ nodes in $P_1$ with gain value $\geq 0$ \label{Refine:filterP2}
	\State \hspace{0.05in} Sort nodes in $L_0$ and $L_1$ with gain value as the key (break ties using node IDs) \label{Refine:sort}
	\State \hspace{0.05in} $l_{min} \leftarrow \min{(|L_0|, |L_1|)}$
	\State \hspace{0.05in} Swap $l_{min}$ nodes with highest gain values between partitions $P_0$ and $P_1$ \textbf{in parallel}	\vspace{0.05in} \label{Refine:swap}
\EndFor	\label{Refine:endFor}
\State Check if the balance criterion is satisfied. Otherwise, move highest gain nodes from the higher weighted partition to the other partition, using a variant of Algorithm~\ref{algInitialLarge}.	\label{Refine:check}
\end{algorithmic}	\label{algRefine}
\end{algorithm}

\subsubsection{Ensuring Determinism.}
In the refinement phase, the only step with potential nondeterminism is Line~\ref{Refine:sort}, in which we create a sorted ordering of the nodes based on their gain values, since there can be multiple nodes with the same gain. BiPart breaks ties between such nodes using their IDs.

\subsection{Tuning parameters}
\label{sec:knobs}

Multilevel hypergraph partitioning algorithms like BiPart have a number
of tuning parameters whose values can affect the quality and runtime of
the partitioning. For BiPart, the three most important tuning parameters are the following.

The first tuning parameter controls the maximum number of levels of coarsening to be performed
before the initial partitioning. Most hypergraph partitioners coarsen the hypergraph until the coarsest hypergraph is very small ({\em e.g.},
PaToH~\cite{patoh} terminates its coarsening phase when the size of the coarsened hypergraph falls below 100). Although one would expect
more coarsening steps to produce a better partitioning, this is not
always the case. For some hypergraphs, we end up with heavily weighted nodes (the weight is the number of merged nodes represented by that node) and processing such nodes in the refinement phase is expensive since they can cause balance problems. In Section~\ref{sec:exp}, we study the performance impact of terminating the coarsening phase at different levels.
The default value used in BiPart is 25.

The second tuning parameter controls the iteration count in the refinement phase.
To obtain the best solution, we can run the refinement until convergence ({\em i.e.}, until the edge cut does not change anymore).
However, this strategy is very slow and thus infeasible for large hypergraphs, which are the focus of this work.
BiPart, by default, uses only 2 refinement iterations.

The final tuning parameter is selecting a matching policy for finding a multi-node matching in a hypergraph.
Table~\ref{t:match_hyperedge} shows the different matching policies available in BiPart.
Some of these policies are based on hyperedge degrees or on the weight of the hyperedge. More policies can be added to the framework by the user. 
The best choice for the policy depends on the structure of the graph, and different policies can result in different partitioning quality as well as different convergence rates. For the experimental results in Section~\ref{sec:exp}, we used LDH, HDH, or RAND, depending on the input hypergraph. 

BiPart exposes these tuning parameters to the application developer but also provides default values for use by novices. Section~\ref{sec:exp} studies the effect of changing these parameters.

\subsection{Parallel Strategy for Multiway Partitioning}
\label{sec:kway}

Multiway partitioning for obtaining $k$ partitions can be performed in two ways: direct partitioning and recursive bisection. In direct partitioning, the hypergraph obtained after coarsening is divided into $k$ partitions and these partitions are refined during the
refinement phase. Recursive bisection uses a divide-and-conquer approach by recursively creating bipartitions until the desired number of partitions is obtained.

In this paper, we present a novel \textit{nested $k$-way} approach for obtaining $k$ partitions. At each level of the divide-and-conquer tree, we apply the three phases
of multilevel partitioning to \emph{all} the subgraphs at that level. Intuitively,
the divide-and-conquer tree is processed level-by-level, and each phase of the
multilevel partitioning algorithm is applied to all the subgraphs at the current level.
Algorithm~\ref{algKWay} presents the pseudocode of our \textit{nested $k$-way} 
approach.

\begin{algorithm}
\small
\caption{Nested $k$-Way Algorithm}
\begin{algorithmic}[1]
\Statex \textbf{Input:} $k$
\For{$level$ $l = 1$ to $\lceil \log k \rceil$ iterations}	\label{kway:for}
	\State \hspace{0.05in} Construct subgraphs $G_1, G_2, \ldots, G_i$ (where $i = 2^{l-1}$) such that $G_j$ contains nodes that are in partition $j$ \label{kway:divide}
	\State\hspace{0.05in} $Coarsen$ $(G_1, G_2, \ldots, G_i)$	\label{kway:coarsen}
	\State\hspace{0.05in} $Partition$ $(G_1, G_2, \ldots, G_i)$	\label{kway:partition}
	\State \hspace{0.05in} $Refine$ $(G_1, G_2, \ldots, G_i)$	\label{kway:refine}
\EndFor	\label{kway:endFor}
\end{algorithmic}	\label{algKWay}
\end{algorithm}

This algorithm allows us to run the parallel loops over the entire edge list of the original hypergraph instead of running them over edge lists for each subgraph separately, which yields a significant reduction of the overall running time.
In Section~\ref{sec:exp-kway}, we present experimental results for obtaining $k$ partitions using this approach.

\section{Experiments}
\label{sec:exp}

We implement BiPart in the Galois 6.0 system, compiled with g++ 8.1 and boost 1.67~\cite{boost}. Galois is a library of data structures and a runtime system that exploits parallelism in irregular graph algorithms expressed in C++~\cite{pingali,nguyen13}.

Table~\ref{bench} describes the 11 hypergraphs that we use in our experiments. The hypergraphs WB, NLPK, Webbase, Sat14, and  RM07R are from the SuiteSparse Matrix Collection~\cite{florida}, Xyce and Circuit1 are netlists from Sandia Laboratories~\cite{zoltan}, Leon is a hypergraph derived from a netlist from the University of Utah, and IBM18 is from the ISPD 98 VLSI Circuit Benchmark Suite.
Random-10M and Random-15M are two hypergraphs that we synthetically generated for the experiments.

All experiments are done on a machine running CentOS 7 with 4 sockets of 14-core Intel Xeon Gold 5120 CPUs at 2.2 GHz, and 187 GB of RAM in which there are 65,536 huge pages, each of which has a size of 2 MB.

We benchmarked BiPart against three third-party partitioners: (i) Zoltan 3.83 (Zoltan is designed to work in a distributed environment; for our experiments, we run Zoltan with MPI in a multi-threaded configuration), (ii) KaHyPar (direct k-way partitioning setting), the state-of-the-art partitioner for high-quality partitioning, and (iii) HYPE, a recent serial, single-level bipartitioner~\cite{hype}. Zoltan and KaHyPar were described in Section~\ref{sec:related_work}. 

The balance ratio for these experiments is 55:45.
Since Zoltan is nondeterministic, the runtime and quality we report is the average of three runs. BiPart numbers are obtained using the configuration discussed in Section~\ref{sec:hypar}.

\begin{table}
\small
 \begin{center}
   \caption{Benchmark Characteristics} \label{bench}
\begin{tabular} {|c|c|c||c|}
\hline
& \multicolumn{2}{c|}{Hypergraph} & Bipartite \\
& \multicolumn{2}{c|}{} & Representation \\
 \hline
 \textbf {Name} &\textbf{Nodes} &\textbf{Hyperedges} & \textbf{Edges} \\
 \hline
\textbf{Random-15M} & $15,000,000$ &  $17,000,000$ &  $280,605,072$\\
\hline
\textbf{Random-10M} & $10,000,000$ &  $10,000,000$ & $115,022,203$ \\
 \hline
   \textbf{WB} & $9,845,725$ &	$6,920,306$ &	$57,156,537$	\\
 \hline
 \textbf{NLPK} & $3,542,400$ & $3,542,400$ & $96,845,792$ \\
\hline
\textbf{Xyce} & $1,945,099$ &	$1,945,099$ &	$9,455,545$ \\
\hline
 \textbf{Circuit1} & $1,886,296$ & $1,886,296$ & $8,875,968$ \\
 \hline
 \textbf{Webbase} & $1,000,005$	& $1,000,005$ & $3,105,536$ \\
\hline
\textbf{Leon} & $1,088,535$ & $800,848$ & $3,105,536$ \\
\hline
\textbf{Sat14} & $13,378,010$ & $521,147$ & $39,203,144$\\
\hline
 \textbf{RM07R} &  $381,689$ &	$381,689$ &	$37,464,962$ \\
 \hline
 \textbf{IBM18} &  $210,613$ &	$201,920$ &	$819,697$ \\
 \hline
 \end{tabular}
\end{center}
 \end{table}

\subsection{Comparison with Other Partitioners}

Table~\ref{tbl:perf} compares BiPart results with those obtained by running  Zoltan, KaHyPar and HYPE. 
BiPart is executed on 14 threads, and 
Zoltan  is executed on 14 processes, while KaHyPar, and HYPE are executed on a single thread since they are serial codes.

KaHyPar produces high-quality partitions but it took more than 1800 seconds to partition large graphs such as Random-10M, Random-15M, webbase, and Sat14. For the hypergraphs that KaHyPar can partition successfully, BiPart is always faster but worse in quality. HYPE runs on all inputs but the execution time and the quality of the partitions are always worse than BiPart.

Zoltan was able to partition all the hypergraphs in our test suite except for the largest hypergraph, Random-15M. 
 For the three largest hypergraphs Random-10M, NLPK and WB, BiPart is roughly 4X faster than Zoltan while producing partitions of comparable quality.
We also compared our results with other hypergraph partitioners, such as PaToH~\cite{patoh} and HMetis~\cite{hmetis}. We observed that the parallel execution time of BiPart is better than HMetis's and PaToH's serial time on large inputs. Since the source code for these partitioners is not available and due to the space constraints, we do not list those results here. We did not compare our results with Parkway since it frequently produces  segfaults.

\begin{table*}
 \small
 \begin{center}
 \caption{Performance of hypergraph partitioners (time is measured in seconds)}  \label{tbl:perf}
    \begin{tabular}{|r|r|r|r|r|r|r|r|r|r|}
        \hline
	    & \multicolumn{2}{c|}{BiPart (14)} & \multicolumn{2}{c|}{Zoltan (14)} & \multicolumn{2}{c|}{HYPE (1)}  & \multicolumn{2}{c|}{KaHyPar (1)}\\ \hline
	    \textbf{Inputs} & \textbf{Time} & \textbf{Edge cut} & \textbf{Time} & \textbf{Edge cut} &  \textbf{Time} & \textbf{Edge cut} & \textbf{Time} & \textbf{Edge cut} \\ \hline
	    \textbf{Random-15M} & \colorbox{green!20}{85.4} & \colorbox{green!20}{13,968,401} &  $-$ & $-$ & $>1,800$ & $15,628,206$ & > 1,800 & $-$ \\ \hline
	    \textbf{Random-10M} & \colorbox{green!20}{35.2} & \colorbox{green!20}{7,588,493} & 133.6 &  8,206,642  & $> 1,800$ & $8,816,800$ & > 1,800 & $-$ \\ \hline
	    \textbf{WB} & \colorbox{green!20}{7.9} & 13,853	& 31.4 & 35,212	& 42.2	& 819,661  & 581.5 & \colorbox{green!20}{11,457} \\ \hline
	    \textbf{NLPK} & \colorbox{green!20}{5.8}	& 98,010 & 	27.6	& 76,987 & 	58.8	& 651,396  & 784.3 & \colorbox{green!20}{59,205} \\ \hline
	    \textbf{Xyce} & \colorbox{green!20}{1.3}	& 1,134 & 4.1 &	1,190	& 11.8 &	549,364	&  412.4 & \colorbox{green!20}{420}	\\	\hline
      \textbf{Circuit1} & \colorbox{green!20}{0.7} & 3,439 & 4.2	& 2,314 & 10.9 &	371,700 & 524.1 & \colorbox{green!20}{2,171}	\\ \hline
	    \textbf{Webbase} & \colorbox{green!20}{0.3}	 &  \colorbox{green!20}{624} &	1.2 &	1,645 &	2.4 &455,492	 & > 1,800 & $-$ \\ \hline
       \textbf{Leon} & \colorbox{green!20}{0.9} &	112 & 5.4	& 81 &	3.8 &32460 &   354.6 & \colorbox{green!20}{59} \\ \hline
       \textbf{Sat14} & \colorbox{green!20}{7.6} & 15,394 & 44.3 & \colorbox{green!20}{5,748} & 61.3 & 524317 & > 1,800  &  $-$ \\ \hline
        \textbf{RM07R} & \colorbox{green!20}{0.8} &	22,350	& 3.9	& 56,296	& 19.1 & 151,570 &  880.0 & \colorbox{green!20}{17,532}	\\ \hline
        \textbf{IBM18} & \colorbox{green!20}{0.2} &	2,669	& 0.4	& 2,462 & 1.0	& 52,779 &453.9  & \colorbox{green!20}{1,915} \\ \hline
    \end{tabular}
 \end{center}
\end{table*}

\begin{center}
\begin{figure}[H]
	\centering
\includegraphics[scale=0.4]{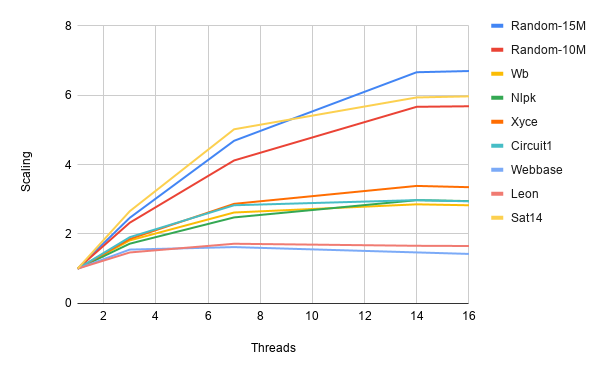}
\caption{Strong scaling of BiPart} \label{fig:scalability}
\end{figure}
\end{center}

\subsection{Scalability}

Figure~\ref{fig:scalability} shows the strong scaling performance of BiPart.
For the largest graphs Random-10M and Random-15M, BiPart scales up to 6X with 14 threads.

Scaling is limited for the smaller hypergraphs like Webbase, Sat14 and Leon since they contain a small number of hyperedges.

\begin{figure}[ht]
    \begin{subfigure}[b]{0.5\textwidth}
      \includegraphics[width=\textwidth]{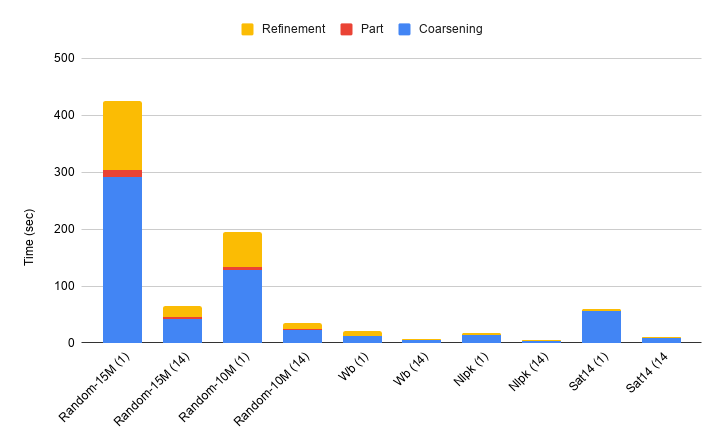}
    \end{subfigure}
    \begin{subfigure}[b]{0.5\textwidth}
      \includegraphics[width=\textwidth]{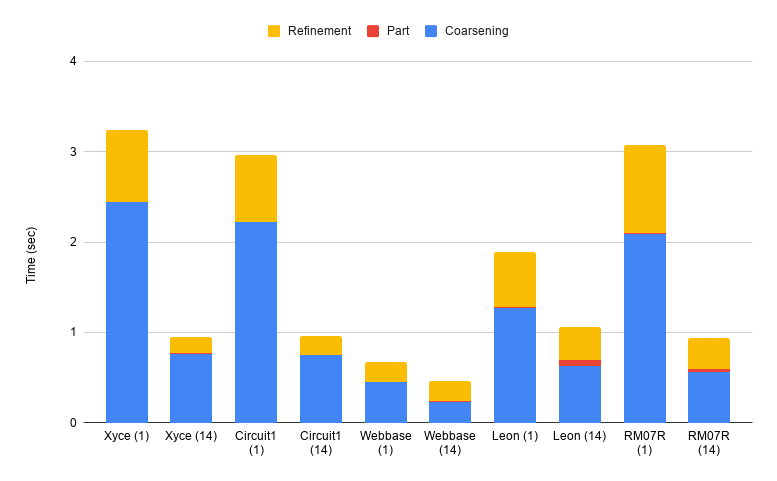}
    \end{subfigure}
      \caption{Runtime breakdown for BiPart on 1 thread and 14 threads.}
      \label{fig:b}
  \end{figure}
Figure~\ref{fig:b} shows the breakdown of the time taken by the three phases in BiPart on 1 and 14 threads, respectively. For both single thread and 14 threads, the coarsening phase takes the majority of the time for all hypergraphs.

The coarsening and refinement phases of BiPart scale similarly.

The end-to-end parallel performance of BiPart can be improved by limiting the
number of levels for the coarsening phase and by a better implementation of the refinement phase. We also see a significant change in the slopes of all the scaling lines when the number of cores is increased from 7 to 8 as well as from 14 to 15. On this machine, each socket has 7 cores so the change in slope arises from NUMA effects. Improving NUMA locality is another avenue for improving the performance of BiPart.

\subsection{Design-Space Exploration of Parameter Space}	\label{sec:pareto}
In this section, we discuss the effect of important tuning parameters on BiPart. The important parameters we explore are the following:
the number of coarsening levels, the number of refinement iterations, and the matching policy.
These parameters are described in detail in Section~\ref{sec:knobs}.

One benefit of having a deterministic system is that we can perform a relatively simple design space exploration to understand how running time
and quality change with parameter settings. In this section, we discuss how the choice of these settings affects the edge cut and running time.

\begin{figure}[ht]
    \begin{subfigure}[b]{0.45\textwidth}
      \includegraphics[width=\textwidth]{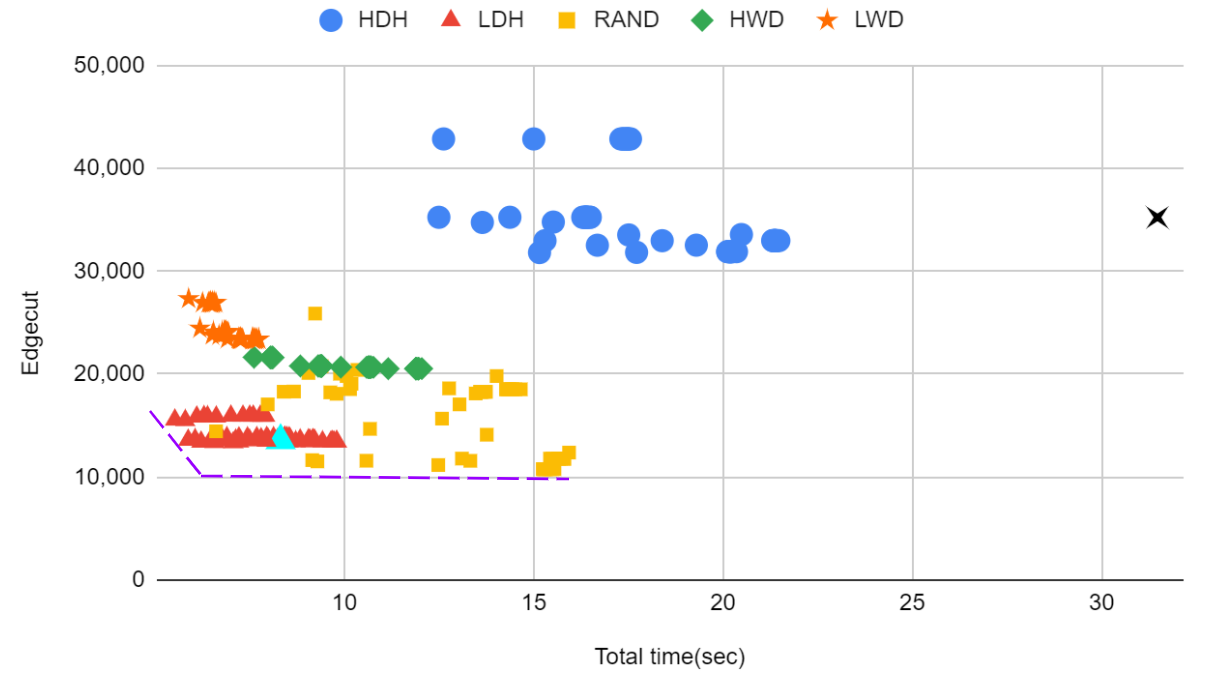}
    \end{subfigure}
    \begin{subfigure}[b]{0.45\textwidth}
      \includegraphics[width=\textwidth]{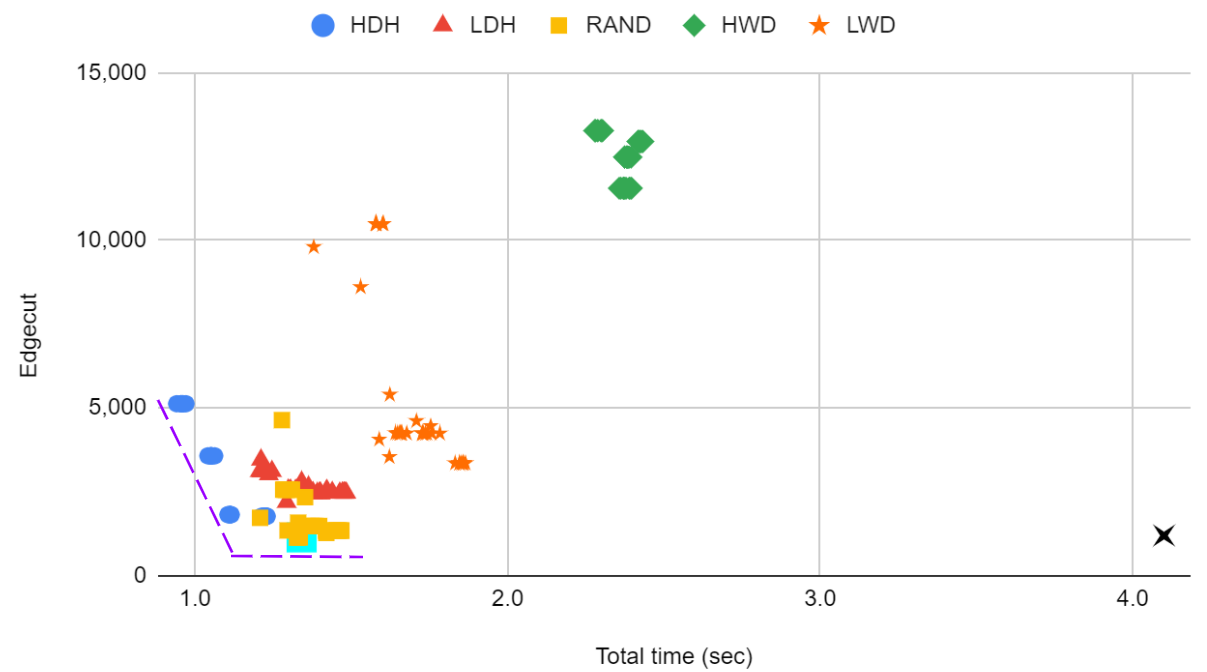}
    \end{subfigure}
  \caption{Design space for various tuning parameters for the two largest hypergraphs, WB (top) and Xyce (bottom); the Pareto frontier is shown for both hypergraphs} \label{fig:par}
  \end{figure}

Figure ~\ref{fig:par} shows a sweep of the parameter space for the two hypergraphs WB and Xyce.
 Points corresponding to different matching policies are given different shapes; for example, 
 triangles represent points for the LDH policy.
While there are many points, we are most interested in those that are on the Pareto frontier. As mentioned in Section~\ref{sec:hypar}, the default settings for BiPart is to
perform coarsening for at most 25 coarsening levels or as much as possible until there is no change in the size of the coarsened graph and to do two iterations of refinement per level.
The BiPart points for this default setting are shown as large circles and triangles (blue in color), and we see that they both lie close to the Pareto frontier. Zoltan points are shown as black \textit{X} marks; for WB, the point is far from the Pareto frontier while for Xyce, the point is on the Pareto frontier but takes much more time for a small improvement in quality.

As for the matching policy for finding a multi-node matching in the coarsening phase, there is no single policy that works best for all inputs. LDH and HDH usually dominate other policies. 
LWD, which has been used in HMetis, does not perform well and does not generate a point on the Pareto frontier, so it should be deprecated.

Table~\ref{tbl:perf-pareto} shows the running time and quality for the default settings, for the settings that give the best quality, and for the settings that give the best running time. The default setting for BiPart is to do two iterations of refinement per level and at most 25 levels of coarsening. For the matching policy, we do not have a fixed matching policy for all graphs but it is a combination of RAND, LD, and HDH. 
For all hypergraphs, the point corresponding to the default setting for BiPart either lies somewhere in between the two extreme points on the Pareto frontier or lies near the Pareto frontier. We also observed that there is no unique parameter setting that guarantees for all hypergraphs that the corresponding point lies on the Pareto frontier.

\begin{table*}
\centering
\small
\caption{Parameter sweep results for BiPart}  \label{tbl:perf-pareto}
	\begin{tabular}{r|r|r|r|r|r|r|}
		\cline{2-7}
		& \multicolumn{2}{c|}{\textbf{Recommended}} & \multicolumn{2}{c|}{\textbf{Best Edge Cut}} & \multicolumn{2}{c|}{\textbf{Best Runtime}} \\ \hline
		\multicolumn{1}{|r|}{\textbf{Graph}}      & \textbf{Time (sec)}   & \textbf{EdgeCut}  & \textbf{Time (sec)}    & \textbf{EdgeCut}   & \textbf{Time (sec)}   & \textbf{EdgeCut}   \\ \hline
		\multicolumn{1}{|r|}{\textbf{Random-15M}}         & 85.4                  & 13,968,401            & 71.4                   & 13,960,994              & 60.7                   & 14,000,612             \\ \hline
		\multicolumn{1}{|r|}{\textbf{Random-10M}}         & 35.2                  & 7,588,493            & 35.3                   & 7,581,745              & 31.4                   & 7,618,589             \\ \hline
		\multicolumn{1}{|r|}{\textbf{WB}}         & 7.9                  & 13,853            & 15.2                   & 10,773              & 6.2                   & 15,904             \\ \hline
		\multicolumn{1}{|r|}{\textbf{NLPK}}       & 5.8                  & 98,010           & 5.8                   & 88,239             & 4.5                   & 121,249            \\ \hline
		\multicolumn{1}{|r|}{\textbf{Xyce}}       & 1.3                   & 1,134              & 1.3                    & 1,134              & 0.9                   & 5,124             \\ \hline
		\multicolumn{1}{|r|}{\textbf{Circuit1}}   & 0.7                   & 3,439             & 1.1                    & 3,408              & 0.5                 & 5,717             \\ \hline
		\multicolumn{1}{|r|}{\textbf{Webbase}}    & 0.3                   & 624               & 0.4                    & 587                & 0.3                  & 622                \\ \hline
		\multicolumn{1}{|r|}{\textbf{Leon}}       & 0.9                   & 112               & 2.1                    & 60                 & 1.5                   & 184                \\ \hline
		\multicolumn{1}{|r|}{\textbf{Sat14}} & 7.6                   & 15,394             & 9.7                    & 13,833              & 2.4                  & 155,325             \\ \hline
		\multicolumn{1}{|r|}{\textbf{RM07R}}       & 0.8                   & 22,350            & 0.9                    & 21,601             & 0.6                  & 30,207             \\ \hline
	\end{tabular}
\end{table*}

\begin{table}
\small
 \begin{center}
 \caption{Performance of BiPart and KaHyPar for k-way partitioning of the IBM18 hypergraph (time in seconds)}  \label{tbl:kway1}
    \begin{tabular}{|r|r|r|r|r|}
        \hline
	    & \multicolumn{2}{c|}{BiPart (14)} & \multicolumn{2}{c|}{KaHyPar (1)} \\ \hline
	    \textbf{$\mathbf{k}$} & \textbf{Time} & \textbf{Edge cut} & \textbf{Time} & \textbf{Edge cut}  \\ \hline
	    \textbf{$\mathbf{2}$} & \colorbox{green!20}{0.2} & 2,385 & 453.9 & \colorbox{green!20}{1,915}  \\ \hline
	    \textbf{$\mathbf{4}$} & \colorbox{green!20}{0.5} & 5,836 & 425.0 & \colorbox{green!20}{2,926}  \\ \hline
	    \textbf{$\mathbf{8}$} & \colorbox{green!20}{1.0} & 11,522 & 288.0 & \colorbox{green!20}{4,822}  \\ \hline
	    \textbf{$\mathbf{16}$} & \colorbox{green!20}{1.6} & 19,116 & 299.5 & \colorbox{green!20}{8,560}  \\ \hline
    \end{tabular}
 \end{center}
\end{table}

\begin{table}
\small
 \begin{center}
 \caption{Performance of BiPart and KaHyPar for k-way partitioning of the WB hypergraph (time in seconds)}  \label{tbl:kway2}
    \begin{tabular}{|r|r|r|r|r|}
        \hline
	    & \multicolumn{2}{c|}{BiPart (14)} & \multicolumn{2}{c|}{KaHyPar (1)} \\ \hline
	    \textbf{$\mathbf{k}$} & \textbf{Time} & \textbf{Edge cut} & \textbf{Time} & \textbf{Edge cut}  \\ \hline
	    \textbf{$\mathbf{2}$} & \colorbox{green!20}{7.9} & 13,853 & 581.5 & \colorbox{green!20}{11,457}  \\ \hline
	    \textbf{$\mathbf{4}$} & \colorbox{green!20}{14.7} & 100,380 & $>$ 1,800 & $-$  \\ \hline
	    \textbf{$\mathbf{8}$} & \colorbox{green!20}{17.5} & 185,079 &  $>$ 1,800 & $-$  \\ \hline
	    \textbf{$\mathbf{16}$} & \colorbox{green!20}{20.0} & 269,144 & $>$ 1,800 & $-$  \\ \hline
    \end{tabular}
 \end{center}
\end{table}

\subsection{Multiway Partitioning Performance}
\label{sec:exp-kway}

\begin{center}
\begin{figure}
\includegraphics[scale=0.4]{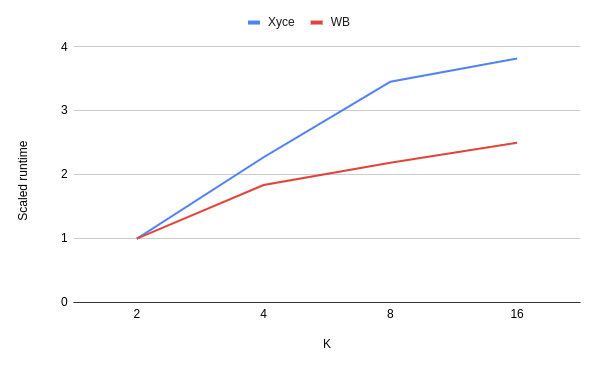}
\caption{BiPart execution time for k-way partitioning} \label{fig:kwayScale}
\end{figure}
\end{center}

Figure~\ref{fig:kwayScale} shows the scaled execution time of BiPart for multiway partitioning of the two hypergraphs Xyce and WB. For both hypergraphs, the execution times are scaled by the time taken to create 2 partitions. If $k$ is the number of partitions to be created, the critical path through the computation increases as $O(log_2(k))$. The experimental results shown in Figure~\ref{fig:kwayScale} follow this trend roughly.

Tables~\ref{tbl:kway1} and \ref{tbl:kway2} show the performance of BiPart and the current state-of-the-art hypergraph partitioner, KaHyPar,
for multiway partitioning of a small graph IBM18 (Table~\ref{tbl:kway1}) and a large graph WB (Table~\ref{tbl:kway2}). We do not compare our results with Zoltan for $k$-way since their result is not deterministic. BiPart is much faster than KaHyPar; for example, KaHyPar times out after 30 minutes when creating 4 partitions of WB (9.8M nodes, 6.9M hyperedges), whereas BiPart can create 16 partitions of this hypergraph in just 20 seconds. However, when KaHyPar terminates in a reasonable time, it produces partitions with a better edge cut (for IBM18, the edge cut is on average 2.5X better).

We conclude that there is a tradeoff between BiPart and KaHyPar in terms of the total
running time and the edge cut quality. As shown in Tables~\ref{tbl:kway1} and \ref{tbl:kway2}, BiPart may be better suited than KaHyPar for creating a large number of partitions of large graphs while maintaining determinism.

\section{Conclusion and Future Work}
\label{sec:conclusions}

We describe BiPart, a fully deterministic parallel hypergraph partitioner, and show that it significantly outperforms KaHyPar, the state-of-the-art hypergraph partitioner, in running time, albeit with lower edge-cut quality, for all inputs in our test suite.
On some large graphs, which BiPart can process in less than a minute, KaHyPar takes over an hour to perform multiway partitioning.

In future work, we want to explore whether we can classify hypergraphs based on features such as the average node degree and the number of connected components to come up with optimal parameter settings and scheduling policies for a given hypergraph. We are also looking into ways to improve NUMA locality for better performance. Extending this work to distributed-memory machines might be useful for very large hypergraphs that do not fit in the memory of a single machine~\cite{cusp}.

\begin{acks}                            
We would like to thank the anonymous reviewers for their insightful feedback, Josh Vekhter for his feedback and help with the figures, Yi-Shan Lu for contributing to the initial paper, and Hochan Lee and Gurbinder Gill for optimizing the BiPart code. This material is based upon work supported by the Defense Advanced Research Projects Agency under award number DARPA HR001117S0054 and NSF grants 1618425, 1705092, and 1725322. Any opinions, findings, and conclusions or recommendations expressed in this material are those of the authors and do not necessarily reflect the views of either the Defense Advanced Research Projects Agency or NSF.  
\end{acks}
\bibliographystyle{ACM-Reference-Format}
\bibliography{paper}


\end{document}